# Enhancing Data Integrity and Traceability in Industry Cyber Physical Systems (ICPS) through Blockchain Technology: A Comprehensive Approach

**Muhammad Imam Hussain** emon.hussain@gmail.com
*State University of New York at Binghamton USA.*

**Md Khairul Islam Bhuiyan** juwel.sust.cse@gmail.com
*Inventive Apps Ltd*

**Shaharier Arafat Sumon** ssumon@g.emporia.edu
*MSIT, Emporia State University USA*

**Shahanaz Akter** shahanazakter@suu.edu
*Southern Utah University, USA*

**Mohammad Ikbal Hossain** mhossai2@g.emporia.edu
*Emporia State University, USA*

**Afroja Akther** afrojabnkdu@gmail.com
*Emporia State University, USA*

**Corresponding Author:** Mohammad Ikbal Hossain

## Abstract

Blockchain technology, originally designed for cryptocurrencies, has emerged as a game-changing innovation with broad potential applications. This paper examines how blockchain can enhance data integrity and traceability in Industry Cyber-Physical Systems (ICPS), which are integral to Industry 4.0. ICPS, which integrate computational and physical elements, are vital in managing critical infrastructures like manufacturing, power grids, and transportation networks. However, these systems face significant challenges related to security, privacy, and reliability. Blockchain, with its inherent features of immutability, transparency, and decentralized consensus, offers a promising solution to these challenges by ensuring data reliability and traceability across ICPS, enhancing transaction transparency, and facilitating secure data sharing.

This study explores several blockchain applications within ICPS, such as supply chain man-agement, quality control, contract management, and secure data sharing. In each of these areas, blockchain has demonstrated the ability to streamline processes, reduce fraud, and im-prove system efficiency. In supply chain management, blockchain enables real-time auditing and ensures compliance. For quality control, it creates tamper-resistant records that build consumer trust. In contract management, smart contracts automate processes, increasing operational efficiency. Additionally, blockchain promotes secure collaboration, which is essential for maintaining the stability and safety of ICPS.

The study highlights the need for further research on the practical implementation of blockchain in ICPS, particularly in addressing challenges like scalability, system integration, and secu-rity vulnerabilities. It also calls for an investigation into the economic and organizational impacts of blockchain within ICPS to assess its long-term feasibility and benefits.

**Keywords:** Blockchain technology, Industrial Cyber-Physical System (ICPS), contact man-agement, data sharing and collaboration, cybersecurity

## 1. INTRODUCTION

Blockchain technology is a revolutionary innovation that has the potential to transform many in-dustries [1]. It is a distributed, decentralized, and secure ledger that can be used to record and track transactions in a transparent and tamper-proof manner [2]. Blockchain technology is the underlying technology behind cryptocurrencies, such as Bitcoin and Ethereum, but it has many other potential applications, such as supply chain management, healthcare, and voting [3, 4]. Blockchain technology, the revolutionary innovation underlying cryptocurrencies like Bitcoin and Ethereum, holds immense potential to transform various industries. Its defining characteristics – security, transparency, decentralization, and efficiency – make it a powerful tool for streamlining processes and enhancing trust across various domains. Blockchain is a way of saving and handling information in a series of connected blocks in computer networks, and it's not limited to just one area of use [2]. Each block can hold different kinds of data, like details about how things are made. This shows how blockchain can be used in systems that manage production [3]. Blockchain's security [1], stems from its distributed ledger architecture, where data is replicated across a network of computers, making it extremely difficult to tamper with or manipulate. This inherent security is particularly valuable in finance and supply chain management, where ensuring data integrity is paramount. Transparency [2], is another hallmark of blockchain technology. Every transaction on a blockchain is publicly viewable, creating an auditable trail of activities. This transparency fosters trust among participants, reducing the need for intermediaries and streamlining processes. Blockchain's decentralized nature implies that no single entity controls the network, eliminating the risk of centralized control or manipulation [3]. This decentralization is crucial for applications like voting systems and decentralized autonomous organizations (DAOs), ensuring fairness and equi-table participation. Blockchain's efficiency lies in its ability to automate processes and eliminate intermediaries. Smart contracts, self-executing contracts embedded in the blockchain, can automate tasks, reducing paperwork, errors, and transaction costs [4]. This efficiency is particularly beneficial in areas like supply chain management and cross-border payments. The potential applications of blockchain technology extend far beyond cryptocurrencies [3]. In supply chain management, blockchain can track the movement of goods and materials, ensuring authenticity, provenance, and efficiency [4]. In healthcare, it can securely store and share medical records, improving patient care and facilitating research [4]. In voting systems, it can enhance security, transparency, and voter confidence [5].

In the era of Industry 4.0, the convergence of physical and cyber worlds has given rise to a new paradigm known as Industrial Cyber-Physical Systems (ICPS) [6]. ICPS are intricate systems that seamlessly integrate computational and physical components, enabling real-time data exchange and decision-making [7]. These systems underpin the operation of critical infrastructure, including manufacturing plants, power grids, and transportation networks [8]. ICPS are characterized by several defining features: deep integration of computational and physical components, real-time data exchange and analysis, autonomous and decentralized control, and adaptive and resilient operations [6]. These features enable ICPS to revolutionize various industries, finding applica-tions in diverse domains such as manufacturing, energy, transportation, and healthcare [7]. De-spite their transformative potential, ICPS face several challenges, including security and privacy, interoperability and standardization, and safety and reliability [8]. Addressing these challenges and harnessing emerging technologies such as artificial intelligence (AI), machine learning (ML), and edge computing will shape the future of critical infrastructure and industrial operations [7]. Blockchain technology, with its inherent properties of immutability, transparency, and distributed consensus, offers a transformative approach to enhancing data integrity and traceability in Industry Cyber Physical Systems (ICPS) [9]. By leveraging blockchain's secure and tamper-proof ledger, ICPS can achieve unprecedented levels of data reliability, traceability, and visibility across the entire ecosystem. Blockchain's distributed ledger technology ensures that data is stored and maintained in a decentralized manner, eliminating the risk of single-point failures and unauthorized modifications [10]. Once data is recorded on the blockchain, it becomes immutable, meaning it cannot be altered or deleted without the consensus of all participants in the network. This immutability provides a robust safeguard against data tampering and manipulation, ensuring the authenticity and integrity of information throughout the ICPS. Blockchain's transparent and auditable nature enables seamless traceability of data throughout the ICPS lifecycle [1, 11]. Every transaction or data modification is recorded on the blockchain, creating an immutable and verifiable record of provenance. This granular traceability allows for the precise tracking of data from its origin to its current state, enabling stakeholders to trace the movement of goods, monitor supply chains, and verify the au-thenticity of products. Blockchain facilitates secure and controlled data sharing among authorized participants in the ICPS [9]. Smart contracts, self-executing programs embedded in the blockchain, can govern data access and permissions, ensuring that only authorized entities can access and utilize sensitive information. This secure and controlled data sharing fosters collaboration and information exchange while safeguarding data privacy and confidentiality. Blockchain's impact on ICPS ex-tends to various applications, including supply chain management, manufacturing and production, industrial asset management, predictive maintenance, and data exchange and collaboration [9]. As blockchain adoption matures, its impact on ICPS will continue to grow, enabling a new era of interconnected, secure, and efficient industrial operations. The primary objective of this research is to explore the potential of blockchain technology in improving data integrity and traceability within Industry Cyber Physical Systems (ICPS). This involves examining how blockchain can provide a secure, transparent, and immutable ledger for data transactions in industrial environments, thereby enhancing overall system reliability and efficiency.

## 2. LITERATURE REVIEW

Industry Cyber-physical system (ICPS), which is a complex network that merges the physical world with computer-based systems. In essence, it includes sensors that collect data from the physical environment, like temperature or speed, and actuators that perform actions in the real world, such as turning on a motor or opening a valve. These components are connected through a network that facilitates the flow of information and commands. Controllers, which can be either hardware or software, play a pivotal role by processing the data from the sensors to make decisions and then sending commands back to the actuators to carry out actions. Human operators may also be part of this system, monitoring and managing it by interacting through the network. Overall, this ICPS framework is fundamental to the functioning of smart devices, autonomous systems, and automated processes in various industries. Industry Cyber-Physical Systems (ICPS) represent a blend of physical processes, networked information technologies, and computational intelligence [12]. This integration of physical and digital worlds is enabling the development of increasingly sophisticated and interconnected systems that can operate with greater autonomy, efficiency, and safety [13]. ICPS are built upon physical processes, which are the fundamental mechanisms by which matter, and energy are transformed [14]. These processes can be as simple as a valve open-ing or closing, or as complex as a chemical reaction or a manufacturing operation. ICPS must accurately model and control these physical processes to achieve their desired outcomes [15]. ICPS relies on networked information technologies to exchange data and information between different components of the system [16]. This includes sensors that collect data about the physical world, actuators that control physical devices, and communication networks that connect these components together. ICPS must effectively manage the flow of information to ensure that the system operates in a coordinated and synchronized manner. ICPS employs computational intelligence to analyze data, make decisions, and adapt to changing conditions [14]. This includes algorithms for machine learning, artificial intelligence, and optimization [15]. ICPS must intelligently process information to make informed decisions that optimize system performance and achieve desired outcomes [16]. The key distinguishing feature of ICPS is the seamless integration of physical processes, networked information technologies, and computational intelligence [12]. This integration enables ICPS to op-erate with a greater degree of autonomy, efficiency, and safety compared to traditional systems [13]. In complex environments like Industry Cyber-Physical Systems (ICPS), ensuring data integrity and maintaining accurate traceability are paramount for maintaining system reliability, safety, and accountability [17]. Traditional methods, such as centralized databases and manual record-keeping, often fall short in addressing the challenges of data security and traceability in these intricate systems [18]. Blockchain technology, with its inherent characteristics of immutability, transparency, and decentralization, emerges as a promising alternative for addressing these challenges [1, 19, 20]. In the context of Industrial Cyber-Physical Systems (ICPS), traditional methods of data management encounter several limitations. Centralized control, as highlighted by Zheng et al. (2018) [21], leads to vulnerability due to single points of failure and potential manipulation. Tian (2016) points out that centralized systems are prone to data tampering and unauthorized modifications, compromising data integrity [17]. These systems also suffer from limited traceability, making it difficult to trace data provenance and establish a clear chain of custody. Furthermore, centralized control often limits transparency and accountability, challenging the verification of data authenticity and the enforce-ment of data integrity policies. To counter these limitations, blockchain technology emerges as a viable alternative. Its key advantages include immutability, ensuring data integrity by preventing alteration or deletion of recorded data [19]. The technology also enhances transparency, with all transactions being publicly visible, allowing stakeholders to verify data authenticity. Decentraliza-tion, another cornerstone of blockchain, eliminates single points of failure, reducing manipulation risks. Additionally, blockchain provides a traceable and auditable trail of data provenance, enabling transparent tracking of data movement and ownership [20].

The applications of blockchain in ICPS are diverse. In supply chain management, blockchain can track goods and materials, ensuring authenticity, provenance, and quality control [17]. It's also beneficial in monitoring manufacturing processes to ensure product quality and regulatory compliance. Asset management is another area where blockchain can track ownership, location, and condition of assets, thus preventing unauthorized access and enhancing management efficiency [18]. Moreover, blockchain facilitates secure data sharing and collaboration within the ICPS ecosystem, and its application in smart contracts automates transactions and enforces agreements, streamlining processes and reducing reliance on intermediaries [21]. ICPS, which merges physical systems with computer-based algorithms, often grapples with issues of data integrity and traceability. The vulnerability of centralized data systems in ICPS is highlighted, especially in the context of cy-berattacks and data breaches [22]. This vulnerability underlines concerns regarding the safety and integrity of critical data. Moreover, Makhdoom et al. (2018) point out the difficulties as-sociated with tracing the origin and authenticity of data in ICPS, primarily due to the complex-ity and lack of transparency in these systems [23]. This problem is exacerbated in sectors like manufacturing and logistics, where data provenance is crucial for quality control and regulatory compliance. Blockchain technology has been identified as a potential solution to these issues. One of the fundamental features of blockchain [24], is its ability to create immutable records, thus ensuring data integrity by preventing tampering and unauthorized alterations. This characteristic is particularly relevant for ICPS, where data accuracy is paramount. Furthermore, the distributed nature of blockchain [25] offers a robust approach against the single-point-of-failure issue prevalent in centralized systems. In terms of traceability, blockchain's transparent and immutable ledger allows for tracking the lineage of data, a feature that Ouaddah et al. suggest could revolutionize data management in ICPS [26]. Blockchain's application extends beyond theoretical discussions. For example, in supply chain management, Wang et al. examine how blockchain can be used to enhance the traceability of products, thus ensuring authenticity and quality [27]. Similarly, in asset management and process monitoring within ICPS, the work of Bahga and Madisetti demonstrates how blockchain can track the lifecycle of products and processes, ensuring compliance and quality assurance [28].

## 3. BLOCKCHAIN TECHNOLOGY

Blockchain technology, in FIGURE 1, is a system of recording information in a way that makes it difficult or impossible to change, hack, or cheat the system. It's a digital ledger of transactions that is duplicated and distributed across the entire network of computer systems on the blockchain. Each block in the chain contains a number of transactions, and every time a new transaction occurs on the blockchain, a record of that transaction is added to every participant's ledger. The decentralized and distributed nature of blockchain makes it highly resistant to unilateral changes or attacks, as the consensus of the entire network is required to validate transactions. The technology is known for its critical role in cryptocurrency systems, like Bitcoin, where it ensures secure and decentral-ized recording of transactions. Blockchain's security comes from its unique use of cryptographic principles, especially using hashes, which are unique codes that secure each block. Changing any single transaction retroactively would require enormous computational resources, making fraud practically infeasible. A key feature of blockchain is the use of consensus mechanisms, such as Proof of Work or Proof of Stake, which are methods used to agree on the ledger's state. They are essential for maintaining uniformity across copies of the ledger and ensuring all participants agree on the verified state of the blockchain without needing a central authority. Blockchain also enables the deployment of smart contracts, which are self-executing contracts with the agreement terms directly written into lines of code. These contracts automatically enforce and execute the terms of an agreement based on its code. The transparency and security of blockchain gives it potential far beyond cryptocurrencies. It's being explored for use in supply chain tracking, secure voting systems, real estate processing, and more, offering a new paradigm for how information is shared and verified in a trustless environment.

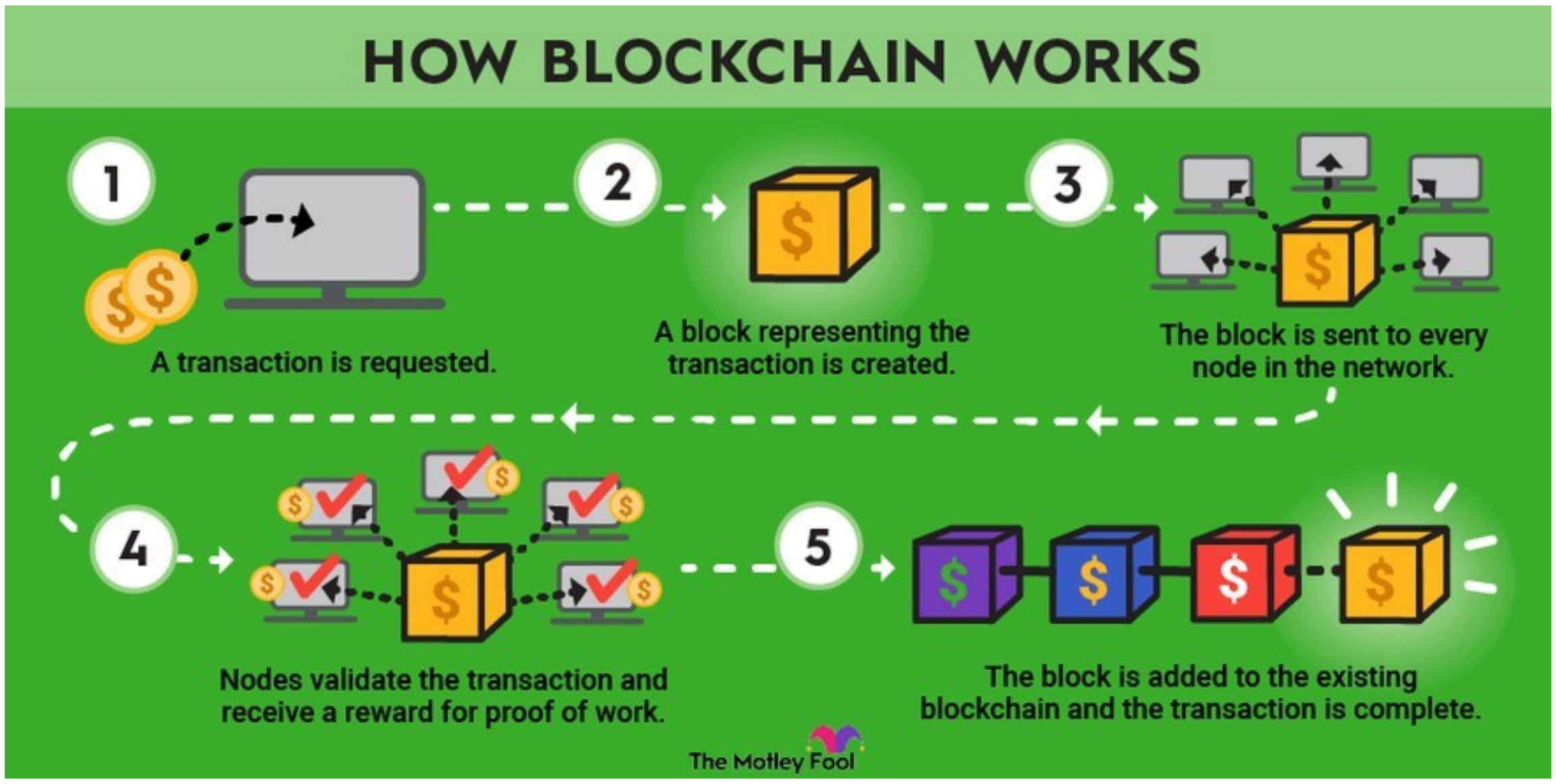

Figure 1: How does Blockchain work?

## 4. EFFECTIVE APPLICATION OF BLOCKCHAIN TO ICPS

The Internet of Critical Physical Systems (ICPS) is a network of interconnected physical and digital systems that are essential for the functioning of modern society. These systems include trans-portation systems, power grids, financial networks, and healthcare systems. Blockchain technology has the potential to revolutionize ICPS by improving the efficiency, security, and transparency of these systems. For example, blockchain can be used to track the movement of goods and materials throughout (TABLE 1) the supply chain, improve the security of financial transactions, and facilitate collaboration among healthcare providers. As blockchain technology continues to develop, we can expect to see even more innovative applications that will improve the ICPS and make our society more resilient and secure.

The ICPS domains outlined in TABLE 1, describe various applications of blockchain technology to enhance efficiency, security, and collaboration across different sectors. In supply chain manage-ment, blockchain enhances the visibility and integrity of the logistics process. For quality control, it secures records and ensures compliance with safety standards. Smart contracts in contract manage-ment automate and streamline execution. Blockchain facilitates secure, transparent data sharing for better collaboration in the ICPS ecosystem and manages identities to prevent unauthorized access. It also optimizes transportation systems, secures e-commerce, manages health records in healthcare, improves financial transactions, and strengthens cybersecurity by protecting against cyberattacks and securing data. Each domain leverages blockchain to solve specific challenges, aiming to foster growth, integration, and security in our increasingly digital world.

Table 1: ICPS Domain

| Key Component | Description | Citation |
|---|---|---|
| Supply Chain Management | Blockchain’s inherent traceability and immutability make it ideal for tracking the movement of goods and materials throughout the supply chain. This can help to improve visibility, reduce fraud, and optimize logistics processes. | [29] |
| Quality Control and Traceability | Blockchain can be used to create a tamper-proof record of quality control data, ensuring that products meet safety and regulatory standards. This can help to improve consumer confidence and reduce the risk of product recalls. | [30] |
| Contract Management | Smart contracts, self-executing contracts that are stored on the blockchain, can automate the execution of contractual terms, reducing the need for manual intervention and paperwork. This can save time and money and improve the efficiency of the contract management process. | [31] |
| Data Sharing and Collaboration | Blockchain can provide a secure and transparent platform for sharing data among different stakeholders in the ICPS ecosystem. This can help to improve collaboration, reduce duplication of effort, and optimize decision-making. | [32] |
| Identity and Access Management | Blockchain creates secure and tamper-proof identities for participants in the ICPS ecosystem. This can help to prevent unauthorized access to sensitive data and systems. | [33] |
| Transportation | Blockchain can improve the efficiency and transparency of transportation systems by tracking the movement of vehicles and goods, managing freight and logistics, and automating payments and settlements. | [34] |
| E-commerce | Blockchain technology offers significant potential for securing e-commerce transactions, safeguarding intellectual property, and reducing fraud, which is particularly beneficial for developing economies in regions like Africa, Asia, and Latin America, where it can greatly enhance economic growth and integration. | [35, 36] |
| Healthcare | Blockchain can secure and manage patient health records, track the movement of pharmaceuticals, and facilitate collaboration among healthcare providers. | [37, 38] |
| Finance | Blockchain improves the efficiency and security of financial transactions, reduce fraud, and promote financial inclusion. | [37] |
| Cybersecurity | Blockchain protects against cyberattacks, secure data, and manage digital identities. | [39, 40] |

### 4.1 Application in Supply Chain Management

Blockchain technology in supply chain management creates a secure, immutable ledger that records every transaction and movement of goods, enhancing data accuracy and security. This allows stakeholders to audit the supply chain in real-time and ensures compliance with various standards. For example, Deloitte discusses the potential for blockchain to reduce supply chain risk, increase visibility, and enhance trust across the supply chain ecosystem. They also highlight how blockchain doesn't displace a company's legacy systems but instead can serve as an add-on enterprise solution [41]. The traditional supply chain, with its multi-layered structure and siloed data, often suffers from inefficiencies, lack of transparency, and vulnerabilities to fraud. Blockchain technology, with its inherent traceability and immutability, offers a transformative solution to these challenges. Every movement of goods and materials is recorded on the blockchain, creating a single source of truth. Authorized participants can access this data in real-time, providing unprecedented visibility into the supply chain. This transparency improves planning, collaboration, and decision-making across the entire network [42]. Consumers increasingly demand knowledge about the origins and ethical practices behind the products they purchase. Blockchain enables detailed provenance track-ing, allowing brands to build trust and loyalty by showcasing their commitment to sustainability and ethical sourcing [43]. Illicit goods and counterfeit products can be easily identified and traced back to their source using blockchain's temper-proof records. This reduces fraud losses and protects brand reputation [44]. Blockchain's secure and transparent payment mechanisms eliminate the risk of fraud and errors in financial transactions within the supply chain. This reduces costs and ensures timely payments for all participants [45]. Automated execution of pre-defined agreements (smart contracts) based on blockchain data eliminates manual intervention and streamlines logistics processes. This reduces lead times, improves efficiency, and minimizes errors [46]. Real-time data on inventory levels and location enables better forecasting and optimization of stock levels. This reduces storage costs, prevents stockouts, and improves product availability [47]. The adoption of blockchain technology in supply chain management offers significant advantages but also presents various challenges. TABLE 2 outlines these challenges and proposes effective strategies for their mitigation.

### 4.2 Application in Quality Control and Traceability

Blockchain technology is rapidly transforming quality control and traceability practices, offering a secure, transparent, and efficient approach to managing product quality. By embracing this tech-nology, businesses can enhance consumer confidence, streamline regulatory compliance, and ul-timately deliver safer, higher-quality products to their customers. Blockchain technology is being increasingly recognized for its role in enhancing quality control and traceability in manufacturing [28]. By using blockchain, a temper-proof record of quality control data is created, which signifi-cantly improves the transparency of inspection results and the detection of defects in manufacturing.

Table 2: Challenges and Mitigation Strategies in the Adoption of Blockchain Technology in Supply Chain Management

| Challenge | Mitigation Strategy |
| --- | --- |
| Integration with Existing Systems | Develop adaptable interface solutions and APIs for seamless connectivity between blockchain platforms and existing systems. Employ a phased implementation strategy to gradually transition from legacy systems to blockchain-based systems. |
| Scalability | Optimize blockchain architecture for high transaction volumes. Utilize scalable consensus mechanisms like Proof of Stake (PoS) or sharding techniques. Invest in research and development to enhance blockchain scalability. |
| Data Privacy and Security Concerns | Implement robust encryption standards. Use private or permissioned blockchains where access is restricted and controlled. Regularly update security protocols and conduct audits to prevent breaches. |
| Technical Complexity and Expertise | Invest in training programs to develop blockchain expertise. Collaborate with blockchain technology experts and service providers. Focus on user-friendly interfaces for ease of use. |
| Regulatory and Compliance Issues | Stay updated with relevant regulations and standards in different regions. Engage with legal experts to ensure compliance. Participate in industry consortia to influence policy making in favor of blockchain technology. |
| High Initial Costs and Investment | Conduct cost-benefit analyses to understand long-term ROI. Seek partnerships and government subsidies where available. Start with pilot projects to demonstrate value before full-scale implementation. |
| Resistance to Change and Adoption | Implement change management strategies. Educate stakeholders about the benefits of blockchain. Garner executive support and develop a culture that embraces technological advancements. |
| Interoperability with Other Systems | Focus on developing blockchain solutions that are compatible with different systems and standards. Collaborate with other blockchain networks to ensure interoperability. |
| Supply Chain Network Participation | Encourage collaboration and participation of all stakeholders in the supply chain. Develop incentives for early adopters and participants. Create a collaborative environment with shared benefits for all parties involved. |
| Reliability and Performance Issues | Conduct extensive testing and quality assurance before deployment. Monitor system performance continuously and have contingency plans in place. Upgrade and maintain the system regularly to ensure optimal performance. |
| User Acceptance and Trust | Build trust through transparency and consistent communication. Showcase successful case studies and pilot projects. Engage users in the development process to align the system with their needs and preferences. |

This is achieved through consensus algorithms like QDT (Quality Decision Tree) which select the most reliable inspection results from multiple blockchain nodes to be added to the ledger. This process helps in detecting defective workpieces at an early stage and ensures that all stakeholders have access to a transparent record of the manufacturing process. These capabilities of blockchain are crucial for ensuring that products meet safety and regulatory standards, which, in turn, can bolster consumer confidence and reduce the risk of product recalls. The BCQC (Blockchain

Quality Controller) framework specifically addresses these needs by fortifying data security and improving the chances of early defect detection [11]. Traditional quality control systems are often hampered by data silos which result in fragmented information, making it difficult to maintain transparency and traceability throughout the manufacturing process. The possibility of record tampering also exists, creating doubts about the authenticity of products and adherence to compliance standards. Moreover, the reliance on manual audits can lead to inefficiencies; these processes are not only slow and susceptible to human error, but they also impact the ability of businesses to respond swiftly and effectively to quality control issues.

Blockchain technology offers a transformative solution to the challenges faced by traditional quality control systems. Firstly, it provides immutable record-keeping by creating a secure, tamper-proof ledger of all quality-related data, ensuring trust and transparency in the supply chain [20]. Secondly, blockchain enhances product traceability by assigning unique digital identities to products, enabling real-time tracking from raw materials to be finished goods. Thirdly, it streamlines audits and compliance by making all data readily available for efficient verification, reducing paperwork, and enhancing responsiveness. Finally, blockchain improves safety and regulatory compliance by implementing and automating quality control protocols directly on the platform, minimizing risks of non-compliance and product recalls. Implementing blockchain technology in quality control and traceability within manufacturing presents unique challenges. TABLE 3 outlines these challenges and provides strategic solutions to address and mitigate them effectively.

### 4.3 Application in Contract Management

The traditional contract management process, often rife with paperwork, manual verification, and human error, can be cumbersome and time-consuming [48]. This inefficiency not only hinders the smooth execution of agreements but also costs businesses money [49]. Enter smart contracts, self-executing agreements stored on the blockchain, poised to revolutionize the way we manage agreements. In essence, smart contracts are computer programs residing on a secure blockchain network. They encapsulate the terms and conditions of an agreement, automatically triggering their execution when predefined criteria are met [50]. This eliminates the need for intermediaries, manual processing, and human intervention, leading to a significantly faster, more secure, and transparent contract management system [51]. Smart contracts offer a range of benefits for contract management, enhancing efficiency, transparency, security, and dispute resolution. Firstly, their au-tomated execution of contractual clauses eliminates the need for manual verification and paperwork, significantly reducing processing time and administrative costs. This automation leads to increased efficiency in managing contractual obligations and deadlines [48]. Secondly, the storage of con-tract terms and transactions on the blockchain ensures that records are tamper-proof and auditable, offering enhanced transparency to all parties involved in the contract. This level of transparency is essential in maintaining trust and accountability [49]. Furthermore, smart contracts reduce the risk of human errors and fraud. The execution of these contracts is governed by predefined condi-tions, which are automatically enforced by the blockchain, minimizing the likelihood of errors or intentional manipulation [51]. This feature is particularly important in complex transactions where the potential for error is high. Additionally, the inherent security features of blockchain technology protect sensitive contract information from unauthorized access, thereby ensuring a high level of data security [48]. Finally, smart contracts simplify the process of dispute resolution. The clear, immutable record of transactions provided by the blockchain makes it easier to resolve disputes [50], as it offers incontrovertible evidence of the terms agreed upon and actions taken. This clarity saves time and resources that would otherwise be spent in lengthy dispute resolution processes. Overall, smart contracts represent a significant advancement in contract management, offering streamlined processes, increased trust, and enhanced security. Smart contracts on the blockchain offer a secure, transparent, and efficient approach to contract management. They automate and streamline the process of executing contractual agreements, enhance trust among parties, and reduce the risk of disputes. However, they also require careful consideration in terms of technological understanding, regulatory compliance, and thorough testing before implementation. Implementing smart contracts in contract management via blockchain technology offers numerous benefits but also presents spe-cific challenges. TABLE 4 outlines five core challenges and their respective mitigation strategies to ensure effective adoption and utilization of smart contracts in contract management.

Table 3: Challenges and Mitigation Strategies in Implementing Blockchain for Quality Control and Traceability

| Challenge | Mitigation Strategy |
|---|---|
| Complexity in Blockchain Integration | Develop comprehensive training programs to build technical expertise. Collaborate with blockchain technology experts for smooth integration. Simplify blockchain interfaces for user-friendliness. |
| Data Privacy Concerns | Implement advanced encryption and privacy-preserving techniques. Use permissioned blockchain networks to control access to sensitive data. |
| Standardization and Compliance Issues | Collaborate with regulatory bodies to develop industry-specific standards. Ensure blockchain solutions are compliant with existing regulations and quality standards. |
| High Initial Investment Costs | Conduct thorough cost-benefit analyses to justify the long-term ROI. Seek partnerships, grants, or subsidies to offset initial costs. Start with pilot projects to demonstrate value and effectiveness. |
| Scalability and Performance | Opt for blockchain platforms that are scalable and can handle high transaction volumes. Regularly update and maintain the blockchain system to ensure optimal performance. |
| Interoperability with Existing Systems | Focus on developing blockchain solutions that can seamlessly integrate with existing quality control systems. Use standard data formats and APIs for interoperability. |
| Reliability of Blockchain Networks | Ensure the blockchain network is robust and capable of handling the specific demands of quality control and traceability. Conduct regular system checks and updates to maintain network health and reliability. |
| Resistance to Technological Change | Implement change management strategies to encourage adoption among stakeholders. Provide clear communication and demonstrations of blockchain benefits to garner support. |
| Ensuring Stakeholder Participation | Engage all relevant stakeholders in the development and implementation process. Foster a collaborative environment with shared benefits to encourage active participation and support. |
| Accuracy and Integrity of Data Entry | Utilize automated data capture technologies to minimize human error. Implement strict protocols for data verification and validation before entry into the blockchain. |
| User Acceptance and Trust | Build trust through transparency and data security assurances. Showcase successful case studies and provide clear, accessible information about how blockchain technology enhances quality control and traceability. |

Table 4: Challenges and Mitigation Strategies for Implementing Smart Contracts in Contract Management

| Challenge | Mitigation Strategy |
|---|---|
| Legal and Regulatory Compliance | Engage with legal experts to ensure smart contracts comply with existing laws and regulations. Adapt contracts to align with legal changes and standards. |
| Technological Complexity | Provide extensive training and resources to develop technical expertise. Collaborate with IT specialists to simplify technology for end-users. |
| Interoperability with Existing Systems | Develop smart contracts that can seamlessly integrate with existing contract management systems. Utilize APIs and standardized protocols for integration. |
| Security Vulnerabilities | Implement robust security protocols and regular audits. Use advanced encryption and access controls to protect sensitive data within smart contracts. |
| User Acceptance and Understanding | Conduct educational workshops and demonstrations to increase understanding and trust in smart contracts. Involve users in the development process. |

## 4.4 Application in Data Sharing and Collaboration

Blockchain technology can significantly enhance data sharing and collaboration in Integrated Cyber-Physical Systems (ICPS), which are increasingly prevalent in industries like healthcare, manu-facturing, and smart city infrastructure. Blockchain's decentralized nature ensures data integrity and security. Since each block in the blockchain is linked to the previous one and is distributed across the network, it is virtually impossible to alter the data without detection. This feature is crucial in ICPS, where data integrity is paramount for system stability and safety [52]. Blockchain provides a transparent and immutable ledger. Every transaction on the blockchain is recorded and can be viewed by all participants, fostering trust among stakeholders. In ICPS, where systems from different sectors may interact, this transparency ensures that all actions are accountable and traceable [53]. Unlike traditional centralized systems, blockchain operates on a distributed ledger technology, eliminating single points of failure and making the system more resilient to attacks and operational disruptions [54]. In ICPS, this can lead to more robust and reliable systems. Blockchain can auto-mate processes through smart contracts – self-executing contracts with the terms of the agreement directly written into code [55]. In ICPS, smart contracts can automatically enforce agreements and actions, streamlining processes and reducing the need for intermediaries. Blockchain can facilitate secure and efficient data sharing among various stakeholders in the ICPS ecosystem [56]. This can lead to improved collaboration, reduced duplication of efforts, and optimized decision-making processes. Blockchain can serve as a standard protocol for different systems and devices in an ICPS to communicate and collaborate, thereby enhancing interoperability [57]. The immutable nature of blockchain records simplifies compliance with regulations [58, 59]. The technology also aids in auditing processes due to its transparent and unalterable record-keeping capabilities. While blockchain technology significantly enhances data sharing and collaboration in Integrated Cyber-Physical Systems (ICPS), it also brings certain challenges (TABLE 5).

Table 5: Challenges and Mitigation Strategies in Implementing Blockchain for Data Sharing and Collaboration in ICPS

| **Challenge** | **Mitigation Strategy** |
|---|---|
| Complex Integration with ICPS | Develop modular and flexible blockchain solutions that can be easily integrated with various ICPS components. Leverage APIs for smooth system integration. |
| Maintaining Data Privacy | Implement permissioned blockchain systems to control data access. Utilize advanced encryption and privacy-preserving technologies for sensitive data. |
| Scalability and Performance | Optimize blockchain architecture for high transaction throughput and reduced latency. Consider scalable consensus mechanisms like Proof of Stake (PoS). |
| Standardization and Interoperability | Collaborate with industry leaders to develop standard protocols for blockchain in ICPS. Ensure compatibility with existing and emerging industry standards. |
| User Training and Acceptance | Conduct comprehensive training programs for stakeholders. Focus on demonstrating the practical benefits and user-friendly aspects of blockchain in ICPS. |

### 4.5 Application in Identity and Access Management

Blockchain technology offers a robust solution for Identity and Access Management (IAM) in Industrial Cyber-Physical Systems (ICPS), addressing critical security and authentication concerns. Blockchain enables the creation of decentralized identities, which are not controlled by any central authority [60]. This approach reduces the risk of identity theft and fraud, as the identity data is distributed across the blockchain network, making it harder to tamper with or forge. Each transaction on a blockchain is recorded in a way that is immutable and time-stamped [61]. This feature is crucial for maintaining an accurate and unforgeable record of who accessed what data and when, providing a clear audit trail for security and compliance purposes. Blockchain can automate access control through smart contracts. These digital contracts can be programmed with specific rules that dictate who can access certain data or systems within the ICPS [62]. Smart contracts execute automatically when their conditions are met, ensuring compliance with access policies without the need for manual intervention. Blockchain uses advanced cryptographic techniques to secure data [58]. These cryptographic methods ensure that only authorized individuals with the correct cryptographic keys can access sensitive information, significantly enhancing the security of the system. In a multi-stakeholder ICPS environment, blockchain can serve as a standard protocol for identity verification across different systems and organizations [61]. This interoperability is essential for seamless and secure interactions within the ecosystem. By using blockchain for iden-tity management, organizations can move away from relying on centralized identity providers, which can be a single point of failure and a target for cyber-attacks [62]. Implementing blockchain technology in Identity and Access Management (IAM) within Industrial Cyber-Physical Systems (ICPS) offers enhanced security but also poses distinct challenges along with strategies to effectively mitigate them (TABLE 6).

Table 6: Challenges and Mitigation Strategies for Implementing Blockchain in Identity and Access Management for ICPS

| Challenge | Mitigation Strategy |
|---|---|
| Complex Integration | Develop integration guidelines and use APIs to ensure smooth interfacing between blockchain IAM solutions and existing ICPS infrastructure. |
| Ensuring Data Privacy | Implement privacy-enhancing technologies such as zero-knowledge proofs and encryption within the blockchain network to protect sensitive identity data. |
| Scalability Concerns | Opt for blockchain solutions designed for high scalability and performance, capable of handling a large number of identity verifications. |
| User Adoption and Training | Provide comprehensive training to users for understanding and efficiently using blockchain-based IAM systems. Highlight benefits to encourage adoption. |
| Regulatory and Compliance Issues | Stay updated with and adhere to data protection laws and industry regulations. Ensure blockchain IAM solutions are compliant with legal standards. |

### 4.6 Application in Transportation

Blockchain technology is increasingly being recognized as a transformative tool for the transporta-tion sector, offering numerous benefits in terms of efficiency, transparency, and automation [63, 64]. Blockchain provides a secure and immutable ledger, ideal for tracking the movement of vehicles and goods [65]. This capability ensures transparency and traceability in the supply chain, allowing stakeholders to verify the authenticity and origin of goods, monitor delivery progress, and optimize routes in real-time. In logistics, blockchain can facilitate better coordination among vari-ous stakeholders, including manufacturers, shipping companies, and retailers [66]. This improved coordination can lead to more efficient inventory management and reduced lead times. Through the use of smart contracts, blockchain can automate complex payment processes and settlements. This automation reduces the need for intermediaries, thereby cutting costs and increasing efficiency [67]. Blockchain's inherent security features, like encryption and decentralization, make it difficult to tamper with data and enhances security in transportation, especially in areas vulnerable to fraud and theft [68]. Blockchain simplifies compliance with various regulations in the transportation industry by providing an immutable and transparent record of all transactions and movements [69]. Blockchain technology can serve as a common platform that ensures interoperability among dif-ferent systems used by various stakeholders in transportation [70]. While blockchain technology holds significant promise for revolutionizing the transportation sector, its implementation comes with specific challenges. TABLE 7 explores challenges and their respective mitigation strategies.

### 4.7 Application in E-commerce

Blockchain technology's application in e-commerce has become a focal point, especially for developing economies in regions like Africa, Asia, and Latin America. Its potential to secure transactions, protect intellectual property, and reduce fraud can substantially contribute to economic growth and integration [36]. Blockchain's decentralized and encrypted nature ensures secure and transparent transactions. It eliminates the need for intermediaries, reducing transaction costs and increasing efficiency. This is crucial in regions where traditional banking infrastructure may be limited [71]. In e-commerce, protecting intellectual property (IP) is vital. Blockchain can be used to create immutable records of IP rights, making it easier to establish ownership and combat counterfeit goods, a significant issue in many developing economies [36, 72]. Blockchain can significantly reduce fraud in e-commerce by providing a transparent ledger where all transactions are recorded and can be audited [73] and this transparency builds trust among consumers and businesses, encouraging more participation in e-commerce. Blockchain simplifies cross-border transactions, which are often complicated by currency exchange and regulatory differences. It enables faster, cheaper, and more secure international trade, essential for developing economies seeking global market integration [74]. The integration of blockchain technology in e-commerce, particularly in developing economies, offers numerous benefits but also faces unique challenges (TABLE 8).

Table 7: Challenges and Mitigation Strategies in Implementing Blockchain in the Transportation Sector

| Challenge | Mitigation Strategy |
|---|---|
| Integration with Existing Transportation Systems | Develop flexible and adaptable blockchain solutions that can integrate with current transportation management systems. Use APIs and standard data formats for seamless integration. |
| Data Security and Privacy | Implement robust encryption and access control mechanisms. Consider using private or consortium blockchains where appropriate to control access to sensitive data. |
| Scalability and Performance | Choose blockchain platforms capable of handling high transaction volumes typical in transportation logistics. Optimize network architecture and use efficient consensus mechanisms to ensure scalability and performance. |
| Regulatory Compliance | Stay informed about transportation regulations and standards. Work closely with legal experts to ensure blockchain solutions comply with national and international laws, especially in cross-border transportation. |
| Stakeholder Collaboration and Adoption | Foster collaboration among all stakeholders, including government bodies, logistic companies, and customers. Conduct educational sessions and workshops to demonstrate the benefits and functionalities of blockchain in transportation. |

### 4.8 Application in Healthcare

Blockchain technology offers transformative opportunities in the healthcare sector, providing solutions for secure management of patient health records, tracking pharmaceuticals, and facilitating collaboration among healthcare providers. Blockchain can revolutionize the way patient health records are managed and shared. By using blockchain, patient data can be stored in a secure, immutable, and decentralized manner. This ensures data integrity, confidentiality, and access control, crucial in handling sensitive health information [75]. Blockchain technology enables the creation of a transparent and unalterable record of the pharmaceutical supply chain, from

manufacturing to

Table 8: Challenges and Mitigation Strategies for Implementing Blockchain in E-commerce

| Challenge | Mitigation Strategy |
|---|---|
| Integration with Existing E-commerce Platforms | Develop adaptable blockchain solutions that can seamlessly interface with existing e-commerce systems. Utilize APIs for integration and data exchange. |
| Ensuring Data Security and Privacy | Implement robust encryption methods and privacy-preserving techniques. Utilize permissioned blockchain networks for sensitive transaction data. |
| Regulatory Compliance and Legal Issues | Work closely with legal experts to ensure compliance with regional and international e-commerce regulations and standards. |
| Scalability and Transaction Speed | Opt for blockchain platforms that can handle high transaction volumes and provide fast processing speeds necessary for e-commerce operations. |
| User Adoption and Trust | Educate consumers and merchants on the benefits and security features of blockchain. Build trust through transparency and reliable transaction records. |

end consumer, crucial in combating counterfeit drugs, ensuring safety, and improving supply chain efficiency [76]. Blockchain enables secure and efficient sharing of patient data among different healthcare providers [77]. This collaborative approach can lead to better patient outcomes, as healthcare professionals can make informed decisions based on comprehensive medical histories. Blockchain technology offers a solution for managing and recording patient consents digitally in a transparent and tamper-proof manner, ensuring compliance with legal and ethical standards for obtaining patient consent [78]. Blockchain can substantially mitigate fraud and abuse in healthcare billing and insurance claims by establishing an immutable and transparent transaction record [78]. Blockchain technology can streamline the management of clinical trials by securely recording and sharing trial data, enhancing the integrity and reliability of clinical research [79]. Blockchain technology holds the potential to revolutionize the healthcare sector, several challenges need to be addressed. TABLE 9 is the key challenges and their corresponding mitigation strategies. These strategies are designed to address the challenges and facilitate the effective implementation of blockchain technology in the healthcare sector, ultimately enhancing the management and sharing of health records, improving patient outcomes, and streamlining healthcare processes.

### 4.9 Application in Finance

Blockchain technology has the potential to revolutionize the finance sector by enhancing efficiency, security, and reducing fraud, thereby promoting financial inclusion. Blockchain's decentralized nature can streamline financial transactions, making them faster and more efficient by eliminating intermediaries. Its cryptographic security also adds an additional layer of protection against cyber threats [80]. The immutable and transparent record-keeping feature of blockchain significantly reduces the potential for fraud in financial transactions. Every transaction is traceable and irreversible, deterring fraudulent activities and ensuring transaction integrity [81]. Blockchain can play a crucial role in enhancing financial inclusion, especially in underbanked or unbanked regions [82]. By enabling secure and low-cost transactions through mobile devices, it can provide access to financial services to those previously excluded. Blockchain technology streamlines and accelerates cross-border financial transactions, overcoming traditional challenges of lengthy processes and high fees, thereby greatly benefiting international trade and remittances [83]. Blockchain technology ensures unmatched transparency in financial operations through its distributed ledger system, where every transaction is recorded, crucial for regulatory compliance and fostering trust among stakeholders [84]. Blockchain facilitates the tokenization of assets, simplifying their trading and management, thereby leading to more efficient markets and broader investment opportunities [85]. Blockchain technology presents significant opportunities for transforming the finance sector, it also faces spe-cific challenges (TABLE 10).

Table 9: Challenges and Mitigation Strategies in Implementing Blockchain in Healthcare

| Challenge | Mitigation Strategy |
|---|---|
| Interoperability with Existing Healthcare Systems | Develop blockchain solutions that can integrate with existing healthcare IT systems using standardized data formats and APIs. This ensures seamless data exchange and compatibility. |
| Data Privacy and Security Concerns | Implement advanced encryption and access control mechanisms within the blockchain framework. Ensure compliance with health data protection regulations like HIPAA to maintain patient confidentiality. |
| Scalability and Performance | Choose blockchain platforms designed for scalability and capable of handling the large volumes of data typical in healthcare. Optimize network architecture to ensure efficient performance in data-intensive applications. |
| Regulatory and Compliance Challenges | Engage with regulatory bodies to ensure that blockchain solutions comply with healthcare regulations and legal requirements. Stay abreast of changes in healthcare laws to adapt blockchain applications accordingly. |
| User Adoption and Training | Conduct comprehensive training for healthcare professionals and administrative staff. Focus on demonstrating the practical benefits of blockchain in healthcare to encourage adoption and ease the transition from traditional systems. |

### 4.10 Application in Cybersecurity

Blockchain technology's application in cybersecurity is a growing area of interest, offering robust solutions for protecting against cyberattacks, securing data, and managing digital identities. Blockchain's decentralized nature makes it inherently resistant to the traditional forms of cyberattacks, such as those targeting centralized databases. The distributed ledger technology ensures that data isn't stored in a single location, making it much harder for hackers to exploit [86]. Blockchain provides a secure and tamper-proof way of storing data. Each block of data is encrypted and linked to the previous block, creating a chain that is nearly impossible to alter without detection [87]. Blockchain can be used to create and manage digital identities in a secure and efficient manner. These digital identities are crucial for online transactions and interactions, providing a reliable and tamper-proof method of verification [88]. While blockchain is transparent, it can also be configured to enhance user privacy and anonymity [89]. This is particularly important in scenarios where per-sonal or sensitive data is involved, such as in healthcare or financial services. Blockchain enables a decentralized approach to cybersecurity by distributing security responsibilities across multiple net-work nodes, reducing the risk of centralized points of failure and enhancing system resilience [90]. Smart contracts on the blockchain can automate various aspects of cybersecurity, such as access control, authentication processes, and compliance enforcement, reducing the potential for human error [91, 92]. Implementing blockchain technology in cybersecurity presents unique challenges, despite its potential to enhance data security and manage digital identities. Incorporating business analytics with blockchain technology in Industrial Cyber-Physical Systems (ICPS) enables data-driven insights that optimize human capital management by enhancing decision-making, workforce efficiency, and operational transparency [93]. TABLE 11 outlines these challenges and proposes effective strategies to address them. These strategies aim to address the challenges and facilitate the effective implementation of blockchain technology in cybersecurity, thereby enhancing the security, resilience, and reliability of digital systems [94].

Table 10: Challenges and Mitigation Strategies for Implementing Blockchain in Finance

| Challenge | Mitigation Strategy |
|---|---|
| Regulatory Compliance and Legal Framework | Work closely with regulatory bodies to ensure blockchain solutions comply with financial laws and regulations. Adapt to evolving regulatory landscapes. |
| Integration with Existing Financial Systems | Develop blockchain solutions capable of integrating with existing financial infrastructure. Utilize APIs and standardized protocols for smooth integration. |
| Scalability and Performance Issues | Opt for blockchain platforms designed for high scalability, capable of handling the large volume of transactions typical in finance. |
| Security and Privacy Concerns | Implement robust security protocols and advanced encryption to protect transaction data. Address privacy concerns, especially in cross-border transactions. |
| User Adoption and Trust | Educate users on the benefits and security features of blockchain in finance. Demonstrate reliability and efficiency through pilot projects and case studies. |

## 5. FUTURE DIRECTIONS AND RESEARCH

This literature review suggests that while blockchain presents a promising solution to the challenges faced by ICPS in terms of data integrity and traceability, further research is needed. Future studies could focus on the practical implementation of blockchain in ICPS, addressing challenges related to scalability, integration with existing systems, and potential security vulnerabilities. Additionally, examining the economic and organizational impacts of blockchain adoption in ICPS could provide insights into its feasibility and long-term benefits.

Table 11: Challenges and Mitigation Strategies for Implementing Blockchain in Cybersecurity

| Challenge | Mitigation Strategy |
|---|---|
| Complex Integration with Existing Security Systems | Develop strategies for seamless integration of blockchain with existing cybersecurity infrastructures, utilizing APIs and custom adaptation methods. |
| Balancing Transparency and Privacy | Implement privacy-enhancing techniques like zero-knowledge proofs within the blockchain to protect sensitive data while maintaining transparency. |
| Scalability and Network Performance | Opt for blockchain architectures that support scalability and high-performance, ensuring the system can handle large-scale cybersecurity applications. |
| Regulatory Compliance and Legal Issues | Stay updated with and adhere to cybersecurity laws and regulations. Ensure blockchain solutions comply with data protection and privacy laws across different jurisdictions. |
| User Training and Adoption | Conduct comprehensive training and awareness programs to educate stakeholders about blockchain's role in cybersecurity. Promote understanding and trust in blockchain-based security solutions. |

## 5.1 Conclusion

The exploration of blockchain technology in enhancing data integrity and traceability within Industry Cyber Physical Systems (ICPS) presents a compelling narrative for the future of industry 4.0. This comprehensive study underscores blockchain's transformative potential, not merely as a foundation for cryptocurrencies but as a pivotal technology for ensuring the security, transparency, and efficiency of ICPS. The integration of blockchain into ICPS addresses critical challenges like data security, traceability, and decentralized control, which are fundamental in managing critical infrastructures such as manufacturing, power grids, and transportation networks. The applica-tions of blockchain in diverse domains—ranging from supply chain management, quality control, contract management, to cybersecurity—demonstrate its versatility and potential to revolutionize traditional systems. Notably, blockchain's ability to create a secure, transparent, and immutable ledger for data transactions significantly enhances overall system reliability and efficiency. This study highlights the various dimensions of blockchain applications, each contributing to a more interconnected, secure, and efficient industrial operation. Whether it's in improving supply chain visibility, ensuring product quality, automating contractual obligations, or facilitating secure data sharing and collaboration, blockchain technology emerges as a robust solution to many of the existing challenges in ICPS. However, the journey towards the widespread adoption of blockchain in ICPS is not without challenges. The research indicates the need for further exploration into the practical aspects of blockchain implementation, such as scalability, integration with existing systems, and addressing potential security vulnerabilities. Moreover, understanding the economic and organizational impacts of blockchain adoption in ICPS is crucial for assessing its feasibility and long-term benefits. Blockchain technology holds the key to unlocking a new era of efficiency, se-curity, and trust in ICPS. Its potential extends far beyond its current applications, promising a future where industrial operations are more resilient, transparent, and efficient. As we advance, continuous research and development in this field will be instrumental in overcoming the existing barriers and fully harnessing the power of blockchain technology in Industry Cyber Physical Systems.